# A Silicon Optical Transistor

Leo T. Varghese,[1,2]* Li Fan,[1,2]* Jian Wang,[1,2]* Fuwan Gan,[3] Ben Niu,[1,2] Yi Xuan,[1,2] Andrew M. Weiner,[1,2] Minghao Qi[1,2,3]†

[1]Birck Nanotechnology Center, [2]School of Electrical and Computer Engineering, Purdue University, 1205 W State St, West Lafayette, IN 47907, USA

[3]State Key Laboratory of Functional Materials for Informatics, Shanghai Institute of Microsystem and Information Technology, Chinese Academy of Sciences, 865 Changning Rd, Shanghai 20050, China

*These authors contributed equally to this work.

†To whom correspondence should be addressed. Email: mqi@purdue.edu. (M.Q.)

**A fundamental road block for all-optical information processing is the difficulty in realizing a silicon optical transistor with the ability to provide optical gain, input output isolation and buffer action. In this work, we demonstrate an all-optical transistor using optical nonlinearity in microrings. By using weak light to control strong light, we observed an On/Off ratio up to 20 dB. It can compensate losses in other optical devices and provide fan-out capability. The device is ultra compact and is compatible with current complementary metal-oxide-semiconductor (CMOS) processing.**

Transistor effect (*1*) is of central importance in information processing. Large logic systems constructed from interconnected transistors are some of the core infrastructures of our society. Meanwhile, high bandwidth long-haul communications are carried by photons. An optical transistor, which allows photons to directly control another stream of information-carrying photons, is an alternative to information processing currently based on electrons and has advantage in integration with high-bandwidth data interconnect, from long-haul to even within electronic chips themselves (*2*). Most proposals or demonstrations (*4-7*) fail to meet a majority of the criteria (*3*) in order to perform logic with light, while the one that did meet (*8*) does not lend itself easy for large scale integration nor is compatible with complementary metal-oxide-semiconductor (CMOS) processing.

Here we propose an architecture for optical transistor that meets the criteria listed in (*3*) and demonstrate its operation in silicon. The basic structure is an asymmetrically coupled add-drop filter (ADF) as shown in Fig. 1A. Analogous to the electrical transistor, the three terminals are labeled as gate, supply and output. The coupling to the ADF is stronger for the gate waveguide than the supply waveguide ($G_1 > G_2$).

With an asymmetrically coupled ADF, Fan *et al*. has demonstrated that the power inside the microring is higher when light is fed from a bus waveguide with a strong coupling than a bus waveguide with a weak coupling for identical input optical power (*9*). When the microring operates in the linear regime (at low optical power), the application of light at the gate terminal does not change the resonance wavelength of the microring. However, at optical power levels high enough to generate nonlinear effects like TPA, FCE and Kerr, heat is generated through phonon vibrations leading to a red-shift in the microring resonance due to thermo-optic effect of silicon (*10*). By operating at optical powers sufficient to reach to red-shift the microrings, we can realize an optical transistor using an asymmetrically coupled microring.

When optical power applied at the supply terminal approaches the ADF in the absence of any optical light at the gate terminal, due to the weak coupling ($G_1$), energy in the microring is not high enough to cause a strong shift in the resonance thereby keeping the microring operating at cold resonance (Fig. 1A). When optical power is added at the gate terminal, due to the strong coupling ($G_2$), energy is built up in the cavity leading to an appreciable red-shift of the microring's resonance keeping it at hot resonance (Fig. 1B). By operating at $\lambda_0$ (Fig. 1C), we can realize a switching effect on the supply power since the output transmission can be turned on or off by controlling the gate power. Furthermore, a transistor effect can be realized since weak light at the gate can control strong light at the supply. Since we are operating at the point where the output changes from high to low when the gate power goes from low to high, the ADF operates in an inverter mode.

In the absence of any power at the gate, optical power at the supply passes the ADF with an insertion loss of only 0.64 dB (Fig. 1C blue spectrum, dashed line) to produce a high output. When the gate is turned on, the output switches to the low state (Fig. 1C red spectrum, dashed line). By switching the gate on and off, an output On/Off ratio of 4.9 dB is observed. Another figure of merit to describe the control of strong light by weak light is the supply to gate ratio (SGR) determined by dividing the supply power applied and the gate power required to switch the output from a high to a low state. With an SGR of ~5.8 dB, the device could be used as a signal re-generation device, analogous to a buffer amplifier in electronics. Hence, this device has fan out capability.

With the basic inverter configuration, the gate wavelength and supply wavelength work at different free spectral range (FSR) to minimize any interference between the two optical sources. However, the device is still cascadable by matching two such ADFs such that the first one's output serves as the second one's gate as shown in <span style="color:red">Fig 1S (supplementary section for cascadability).</span>

The optical inverter has a static power transfer characteristic (PTC) similar to the voltage transfer characteristic (VTC) of the CMOS inverter, as shown in Fig 1D. In VTC, the switching threshold voltage is defined as the point where the input voltage equals the output voltage (*11*). Unlike the CMOS inverter, our optical device's threshold power is defined as the gate power at which the output switches from one state to the next. As indicated by the PTC, below a gate power of $P_L$, the output remains high. Above a gate power of $P_H$, the output is low. There is a sharp transition between $P_L$ and $P_H$ where the output is indeterminate as indicated by the large error bar. This is because there is insufficient power at the gate to fully turn on the ADF. In this power range, the ADF state is not stable and fluctuates between two states. The transition point $P_M$ is defined as the device threshold power. A look up table (LUT) is tabulated based on the operation of the PTC (Table 1).

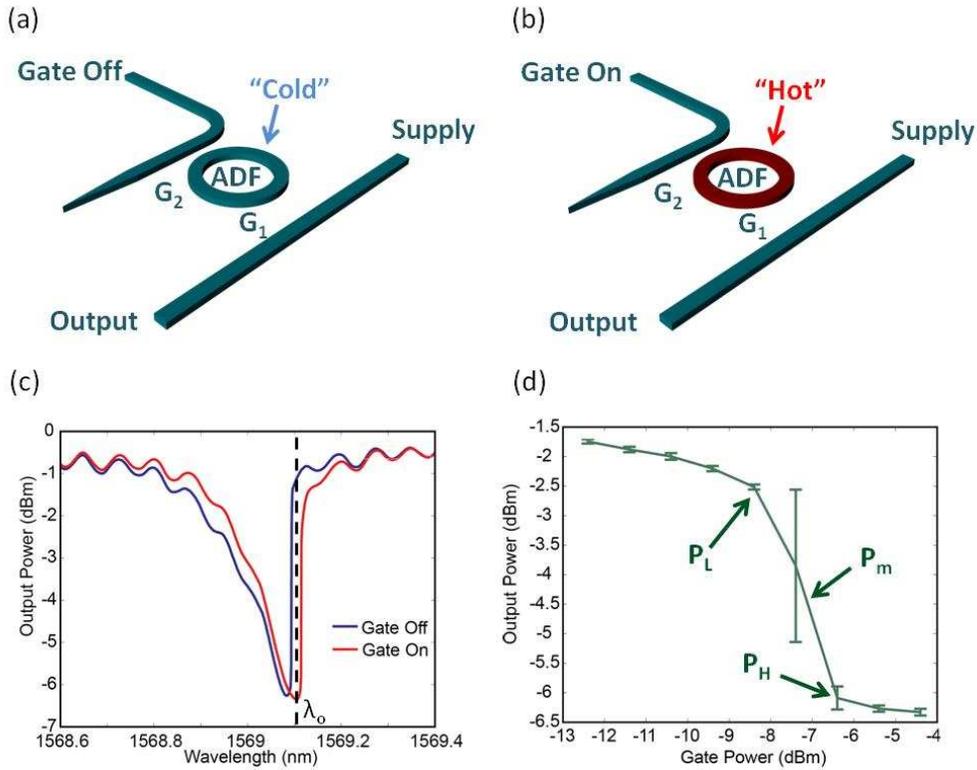

Figure 1 Schematic and operation of the optical transistor. (A-B) The schematic of the ADF with asymmetrically coupled ring with gaps $G_1$ (450nm) and $G_2$ (400nm). It shows the gate, supply and output ports of the system. While (A) shows the ADF is cold when the gate is off, (B) shows the ADF is hot when the gate is turned on. (C) A continuous scan of the ADF response with gate signal off and on. The dashed line shows the operating point of the optical transistor at 1569.105 nm with a supply power of ~-0.55 dBm and a control power of ~-6.4 dBm at 1552.385 nm to realize a SGR of ~5.85 dB. (D) The static power transfer characteristic of the optical inverter at a supply power of ~-0.55 dBm with the solid line going through the mean and the error bar indicating the standard deviation of the points. $P_L$ is the low power at the gate at which the output power is high, $P_M$ is defined as the gate threshold power at which the output state is undetermined and $P_H$ is the high power at the gate when the output power is low.

| Gate Power (P$_G$) | Input Logic | Output Logic |
|:---:|:---:|:---:|
| P$_G$<P$_L$ | LOW | HIGH |
| P$_L$<P$_G$<P$_H$ | N/A | Indeterminate |
| P$_G$>P$_H$ | HIGH | LOW |

Table 1 Operation point of the optical inverter and logic states of the input (gate) and output.

Although an optical inverter can be realized with an ADF configuration alone, it suffers from poor On/Off ratio. Optical transistors have unique requirements that are different from their electronic counterparts – namely – the OFF state should have extremely low transmission. This is because the sum of the voltage in electronics does not add up, while in optics, small powers may add up. If the OFF state has -3dB transmission, then adding two OFF state transistors will become an ON state. This means "0+0=1". In order for standard logic addition or multiplication to function, the OFF state must be much lower than the "threshold" intensity, preferably around – 20 dB or 1%. This would allow the AND operation for reasonable amount of input, i.e. 10~20 to remain logically the value of "0".

To improve this figure of merit, we cascaded a notch filter (NF), consisting of a 5 micrometer ring critically coupled to the bus waveguide, after the ADF on the supply side (Fig. 2A&B). A titanium/gold micro-heater (not shown) is placed to the side of the NF to thermally tune the microring to operate in-conjunction with the ADF.

In the linear regime, the act of switching the gate on and off does not change the output of the ADF+NF device, similar to the ADF only configuration. However, when the powers are high enough to induce non-linearity in the microrings, an optical inverter can be realized by thermally tuning the NF to operate its cold resonance slightly to the right of the ADF's cold resonance.

When the gate is off (Fig. 2A), the supply power passes the ADF with little attenuation due to the large gap ($G_1$) and reaches the NF, causing the it to red-shift significantly, allowing the optical light to pass, albeit with some insertion loss (~4.8 dB) to realize a high output (Fig. 2C blue spectrum, dashed line). When the gate is switched on (Fig 2B), the ADF red-shifts due to the small gap ($G_2$) leading to the supply power being attenuated. This attenuated power reaches the ADF cannot appreciably red-shift the NF producing a significant dip at cold resonance which greatly attenuates the optical signal reaching the output, thereby realizing a low output (Fig. 2C red spectrum, dashed line). By cascading an NF to the ADF, an On/Off ratio of 19.26 dB is realized. Although the insertion loss is relatively high, with an SGR of ~8.05 dB, a moderate net gain of 3.25 dB is observed.

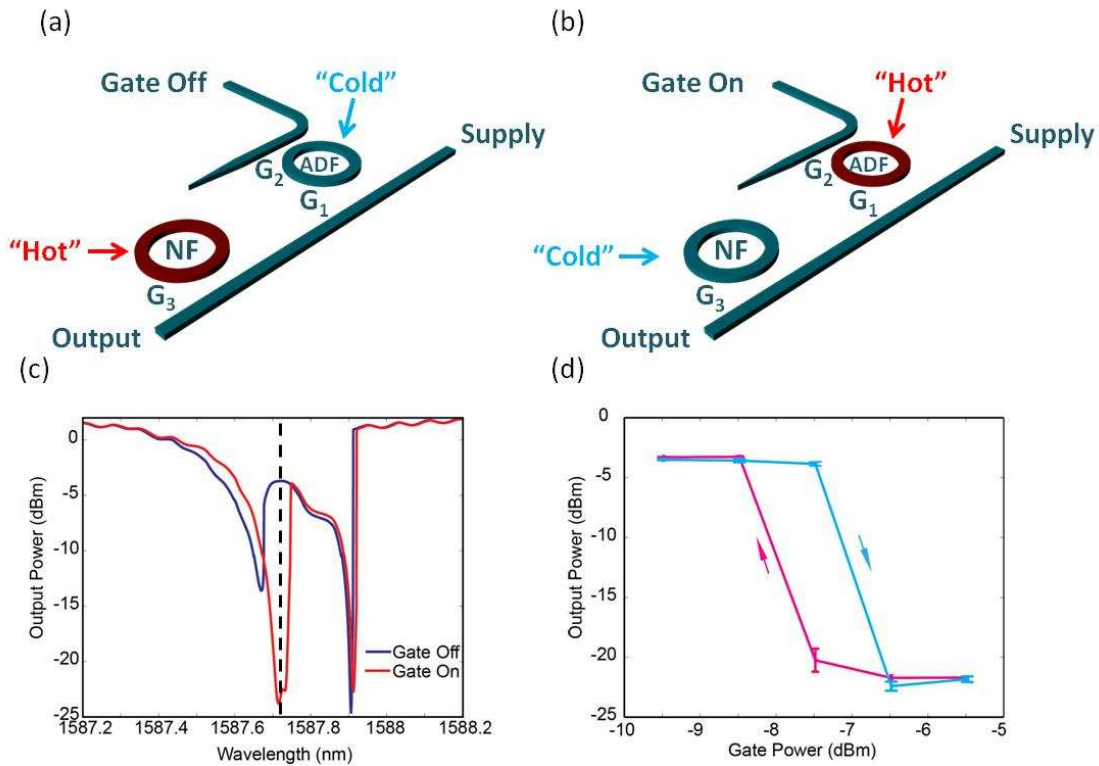

Figure 2 Schematic and operation of the optical transistor with NF cascaded to the ADF. (A-B) The schematic of the configuration with asymmetric gaps $G_1$ (450nm), $G_2$ (400nm) for ADF and gap $G_3$ (380nm) for the NF. It shows the gate, supply and output ports of the system. (A) When the gate is off, the ADF is at cold resonance while the NF is at hot resonance (in red) due to the supply power. (B) When the gate is turned on, the ADF is hot while the NF remains at cold resonance. (C) A continuous scan of the hybrid optical transistor response with gate off (blue) and on (red). The dashed line shows the operating point at 1587.715 nm with a supply power of ~1.6 dBm and a control power of ~-6.45 dBm at 1570.26 nm. (D) The PTC hysteresis plot of the hybrid optical inverter at a supply power of ~1.6 dBm.

In Fig. 2D, a PTC hysteresis plot is observed by sweeping the optical light at the gate up and down rather than switching it off and on. Due the bistability of the ADF microring, the output between -8 and -6 dBm is dependent on the direction of the sweep of the gate signal. This could be used as anti-jitter control and would allow the cleaning up of lousy signals and produce clear low or high outputs. By operating beyond the hysteresis region, we can switch the hybrid optical transistor on and off with a small change in the gate power (Fig. 3).

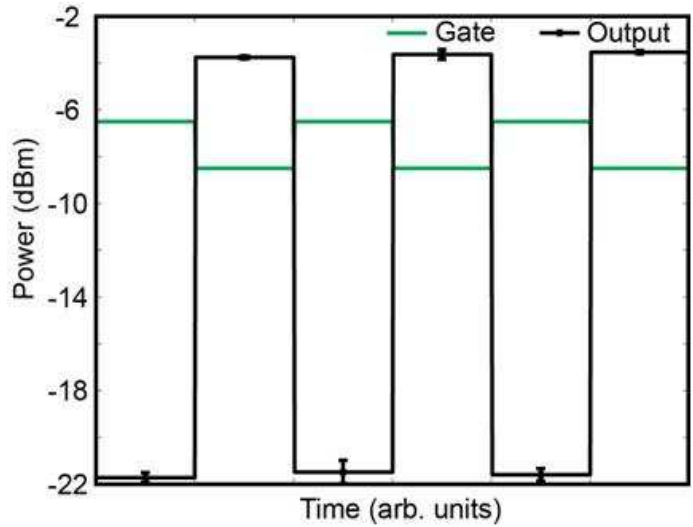

Figure 3 Switching of the hybrid optical transistor. The supply power is at ~2.75 dBm operating at 1587.706 nm. When the gate power (green) is switched between ~-6.5 dBm and ~-8.5 dBm (a difference of ~2 dB), we observed an output between ~-22 dBm to ~-4 dBm (a difference of 18 dB) with error bars for the output power.

Another requirement important for optical integration is the power gain. This differentiates a transistor from a modulator, where no power or signal gain is required. Power gain is especially important for several reasons. First, the coupling of signals into the device typically involves loss (insertion loss). If no optical power gain is achieved, first the logic operation cannot propagate down the chain, also it does not allow power monitoring and self-adjustment (feedback, diagnosis), etc. Another important feature is fan-out, i.e. an output should be able to drive at least two similar gates, which means a power gain of at least 3 dB.

Although the optical transistor is based on resonance, the bandwidth demonstrated in this work is enough for data rates up to 10 GHz and possibly higher. Bit error rate testing was also performed with an external modulator at 10 GHz which clearly shows that the optical transistor performs close to the back-to-back experiments when the gate is off and does not allow light to pass when the gate is turned on. To demonstrate this, we used an on-chip modulator fabricated in a CMOS foundry [REF] and cascaded to our optical transistor. High speed packets pass successfully with an open eye for a 5 GHz $2^{31}$ pseudorandom bit sequence (PRBS) when the gate is switched off (Fig. 4A). When the gate is turned on, the packets are distorted due to inter-symbol interference and power attenuation.

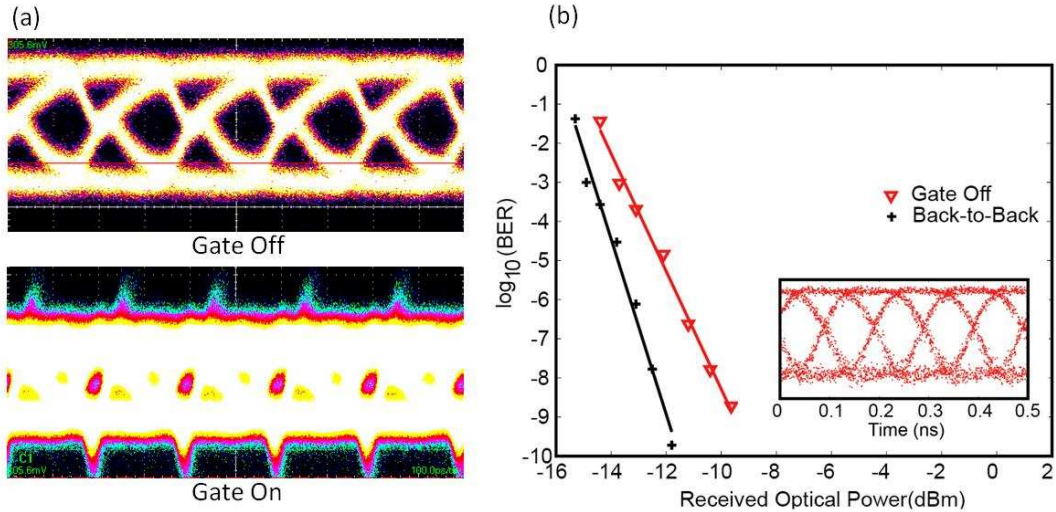

Figure 4 High speed packets through the hybrid optical transistor. (a) A 5GHz eye for an on-chip Mach-Zehnder modulator passed through the hybrid optical transistor when the gate is switched on or off. (b) Bit

error rate testing of the hybrid optical transistor at 10GHz using an external modulator with the inset showing a clear eye when the gate is switched off.

The device uses the asymmetric coupling of a microring to demonstrate the control of strong light by a weak light. With the addition of a notch filter operating at critical coupling, we demonstrated high on/off ratio which may be crucial for optical information processing. The realization of a silicon optical transistor with CMOS technology allows us to come one step closer to an all optical computing.

**Materials and Methods**

Sample Fabrication

We fabricated the device on a silicon-on-insulator wafer (from SOITEC) with a 250 nm-thick top silicon layer and 3 μm of buried oxide. The device has a rectangular cross-section of 250 nm in thickness and 500 nm in width, which supports a low-loss single-mode quasi-transverse magnetic (TM) mode. It was patterned with high resolution electron-beam lithography (Vistec VB6) which has a beam step size of 2 nm. First, the gratings were patterned and etched to form the vertical grating couplers (VGC). Next the optical transistor was patterned and transferred into silicon. Both etching steps were performed through reactive-ion etching using $Cl_2$/Ar. Titanium/gold micro-heaters with 800 Ω resistances were evaporated next to the NF only. SU8 resist of ~3μm was written over the gratings on the VGC to efficiently couple TM mode light in and out of the device. No over-cladding was deposited over the rest of the device.

Experimental Setup and Measurement Details

　　To efficiently couple light into the device's VGC, a fiber array consisting of 4 single mode fibers glued together, 250μm apart, with a fixed angle of 8° is mounted on a 5 axis nanopositioned stage. One continuous-wave tunable laser source was fed to the supply terminal and a second continuous-wave tunable laser source was fed to the gate terminal. The output terminal of the optical transistor was connected to an optical power meter.

　　To tune the resonance of the NF, the micro-heater to the side of the NF was heated using a constant voltage source so that the resonance of the NF at low power is just to the right of the ADF resonance.

　　Continuous measurements are undertaken by sweeping the wavelength of the tunable laser from small wavelengths to large wavelengths with a constant power. Point measurements are done by applying a constant power at the operating wavelength and noting the output power after the laser is turned on. Each data point is obtained by averaging the result obtained from 5 repeated measurements. The operating wavelengths of the gate terminal are chosen to be on a different FSR from the supply terminal. The power transfer characteristic (PTC) plot is obtained by noting the optical power at the output terminal as the power to the gate is switched on and off. To avoid any hysteresis due to bistability, the power is changed only when the gate is turned off. Throughout this measurement, the supply power and wavelength is kept constant. The PTC hysteresis plot is obtained in the same fashion with one difference: the gate power is never switched on/off but rather the optical power is changed.

Author Contributions

L.T.V. and L.F. fabricated and characterized the devices. J.W. performed simulation. F.G. designed the on-chip modulator. B.N. helped in the design. Y.X. helped in the fabrication. M.Q. conceived the idea and supervised the investigation. L.T.V., L.F and M.Q. wrote the manuscript. All discussed the results and commented on the manuscript.